\documentclass[a4paper,11pt]{article}
\usepackage{epsfig,array,multirow,axodraw}  
\usepackage{amsbsy,amssymb}
\newcommand{\eqn}[1]{Eq.(\ref{#1})}
 
\newcommand{\bsm}{\begin{small}} 
\newcommand{\esm}{\end{small}} 
\newcommand{\bc}{\begin{center}} 
\newcommand{\ec}{\end{center}} 
\newcommand{\bl}{\begin{large}} 
\newcommand{\el}{\end{large}}
\newcommand{\bL}{\begin{Large}} 
\newcommand{\eL}{\end{Large}} 
\newcommand{\bh}{\begin{huge}} 
\newcommand{\eh}{\end{huge}} 
\newcommand{\ben}{\begin{enumerate}} 
\newcommand{\een}{\end{enumerate}} 
\newcommand{\bit}{\begin{itemize}} 
\newcommand{\eit}{\end{itemize}} 
\newcommand{\bq}{\begin{equation}} 
\newcommand{\eq}{\end{equation}} 
\newcommand{\bqa}{\begin{eqnarray}} 
\newcommand{\eqa}{\end{eqnarray}} 
\newcommand{\nn}{\nonumber}

\newcommand{ \feyn } {\ensuremath { \put(-9,0){ / }}}
\newcommand{ \cfeyn } {\ensuremath { \put(-10,0){ / }}}
\def\demo{$\Delta\eta\mu \acute{o} \kappa \varrho \iota \tau o \varsigma$}

\begin{document}
\title{ {\bf  MULTI-JET PRODUCTION IN HADRON COLLISIONS} }
\author{ Petros D. Draggiotis\thanks{petros@sci.kun.nl}\\
        {\it University of Nijmegen, Nijmegen, The Netherlands}, \\
        {\it Institute of Nuclear Physics, NCSR \demo, 15310 Athens, Greece} 
       \and
         Ronald H. P. Kleiss\thanks{kleiss@sci.kun.nl}\\
        {\it University of Nijmegen, Nijmegen, The Netherlands}
       \and
        Costas G. Papadopoulos\thanks{Costas.Papadopoulos@cern.ch}\\
        {\it Institute of Nuclear Physics, NCSR \demo, 15310 Athens, Greece}} 
\maketitle
\begin{abstract}
The advent of high-energy hadron colliders necessitates efficient and
accurate computation of multi-jet production processes, both as QCD processes
in their own right and as backgrounds for other physics. 
The algorithm that performs these tasks and a brief numerical study of multi-jet processes
are presented.
\end{abstract}
\section{Introduction}
With the recommissioning of the Tevatron, and the forseeable commencement
of physics at the LHC, the need for fast and accurate QCD calculations
is now larger than ever. In this paper, we describe our efforts to arrive
at such calculations. To set the stage: we have in mind the computation
of QCD cross sections with many observed jets that, as usual, are
modelled by assuming that each jet comes from a single fragmenting parton.
With the large amount of energy available, the number $n$ of jets can easily
be as large as 8, thus requiring the computation of amplitudes with 10 or
even more external legs.

Although in principle straightforward enough, the usual techniques of
evaluating Feynman diagrams and integrating the resulting cross section by
Monte Carlo are in practice hampered by the computational complexity of
the problem. The following obstacles can be recognized:
\begin{enumerate}
\item
 The flavours of the initial partons are never detected, and in
 most cases (barring, say, $b$ tagging) neither are the flavours of the
 final-state partons. In what follows we take all quarks
 ($u$, $d$, $s$, $c$, and $b$) to be  essentially massless, although a version
 of the code with massive fermions exists. In the case of flavour integration,
 to be discussed later on, we always treat quarks as massless.
 This means that in any given jet configuration there are usually very many
 contributing processes. Even enumerating these is a nontrivial task,
 and calculating each individual process cross section is even more so \cite{DK}.
\item 
 In addition to flavour, neither the colour nor the spin of any parton 
 is observed. For an amplitude with $p$ quark partons and $q$ gluons this
 implies that in principle $(6)^p(16)^q$ contributions have to be added.
 It is true that, in particular, many colour configurations lead to
 zero amplitudes, but figuring out precisely which these are is very hard.
\item
 Each individual amplitude, with specified flavours, colours and spins,
 contains very many Feynman diagrams. In Appendix B we give a recipe for
 determining the precise number of graphs, at the tree level \cite{diagramcountingR&P}.
 Typical results
 are that the process $gg\to 8g$ is described by 10,525,900 diagrams, and
 the process $gg\to 2g3u3\bar{u}$ by 946,050 diagrams. Inclusion of
 loop corrections worsens this dramatically, of course.
\item
 Each amplitude peaks in complicated ways inside the momentum phase space.
 Straightforward integration is therefore impractical, and one has to
 search for efficient mappings to do importance sampling in a multi-particle
 phase space.
\end{enumerate}
All these difficulties are addressed in this paper, and here we describe
our solutions, in reverse order.
\begin{enumerate}
\item[4.]
 The peaking structure of the amplitude is dealt with by our phase-space
 generating algorithm {\tt SARGE} \cite{HKD,HaKl}. 
 This algorithm is tailored to
 the generation of so-called antenna structures. Let $p_1,\ldots,p_n$ be
 the momenta of the $n$ partons involved. 
 The peaking behaviour of the cross section for the purely
 gluonic process involving $n$ gluons, $gg \to (n-2)g$ is then dominated by the following antenna
 structure:
 $$
 \left[\vphantom{{A\over a}}
 (p_1p_2)(p_2p_3)(p_3p_4)\cdots(p_{n-1}p_n)(p_np_1)\right]^{-1}
 $$
 and any permutation of labels. Since processes involving quarks do not show
 other dominant peaking behaviour than the above ones, we can cover the
 $n$-particle momentum phase space with good efficiency. More details about
 {\tt SARGE} can be found in \cite{vHam}
\item[3.]
 Over the last years new algorithms, along with their implementations, 
 for computing the tree-order scattering amplitude have been 
 proposed~\cite{CarMor,DragPap,helac}. 
 These do not involve calculation of individual diagrams, but rather
 reorganize, in a systematic way, the various off-shell subamplitudes in such a way that
 as little of the computation as possible is repeated. The improvement in
 computational efficiency of these algorithms is nevertheless dramatic, down from about
 $n!$ to something like $3^n$. 
 In the algorithm suggested originally in reference~\cite{CarMor} the scattering amplitude
 was computed through a set of recursive equations 
 derived from the effective action as a function of the classical
 fields. These classical equations represent nothing but the
 tree-order Schwinger-Dyson equations, a fact that was already emphasized in subsequent
 approaches~\cite{helac} and will be illustrated below for the special case of
 QCD. In fact, recursive techniques have already been used in the past 
 to compute multi-gluon amplitudes\cite{Giele}. 
\item[2.]
 The usual spin and colour summation is replaced by a Monte Carlo integration.
 In the purely gluonic case \cite{DragPap} this has been shown to work
 {\em rather} well. It must be realized that since we replace the more usual sum
 over discrete colour and spin states by a continuous (Monte Carlo) integration,
 the variance of the cross section can indeed be expected to be smaller.
\item[1.]
 We tackle the flavour combinatorics in the same spirit: we
 assign to each quark and antiquark a flavour quantum number, that is,
 we extend the definition of the external legs by a direct product with
 a flavour vector in the space of the 5 available flavours. By taking
 these vectors randomly we can perform a sum over flavours by Monte Carlo
 as well, and the only computational difference between the case of
 one single flavour and that of $f$ flavours is simply a factor $f$, whereas
 the discrete-flavour-sum combinatorics would lead to a much bigger
 loss in speed. The possibility of coherent superpositions of different
 flavours might seem awkward but is in fact quite natural since all quarks
 are treated as massless and therefore the distinction between flavours
 is to a large extent arbitrary. The only place where flavours are {\em not\/}
 treated on an equal footing is in the structure functions that describe
 the difference in probability of picking out different flavours: by
 a judicious weighting of the flavour vectors  for incoming quarks and
 antiquarks we can handle this as well.
\end{enumerate}
Considering the Monte Carlo treatment of spin, colour and flavour, the
essential point is to realize that we are entitled to use {\em any\/}
representation for the corresponding information, as long as we end up
with the correct sums {\em on the average\/}.

Before finishing this introduction we want to mention that, so far, we
can calculate partonic cross sections. These are typically good for
global quantities like total cross sections, $p_T$ distributions and the
like. A fuller treatment involves coupling the generated events to
fragmentation programs like {\tt HERWIG} \cite{herwig}. These programs
typically require additional information on the `spanning of the colour
string'. In \cite{manganoetal} a way to do this for processes with zero or
one quark lines is described; in \cite{colourstrings} we shall indicate how
the present program can be adapted in a similar manner.
\section{The algorithm for jet production}

\subsection{The scattering amplitude}
Our starting point are the  Dyson-Schwinger (DS) equations, which give
recursively the $n$-point Green's functions. 
These equations hold all the information for the fields and their interactions
for any number of external legs and to all orders in perturbation theory. We restrict
ourselves to tree level calculations, so we discard ghost fields.
Also we consider all quarks as massless. The couplings between fields, relevant
for QCD, are shown below.

\begin{center}
\SetScale{0.6}
\fbox{
\begin{picture}(250,230)(-30,-20)
\Gluon(0,10)(25,35){5}{4}
\Gluon(0,60)(25,35){5}{4}
\Gluon(25,35)(50,60){5}{4}
\Gluon(25,35)(50,10){5}{4}
\Vertex(25,35){2}
\Text(0,45)[]{$a, \alpha, k_1$}
\Text(0,0.8)[]{$d, \delta, k_4$}
\Text(40,45)[]{$b, \beta, k_2$}
\Text(40,0.8)[]{$c, \gamma, k_3$}
\Text(120,18)[]{$-g^2 G_{abcd}^{\alpha \beta \gamma \delta} $}
\Gluon(0,120)(30,150){5}{4}
\Gluon(30,150)(60,120){5}{4}
\Gluon(30,150)(30,190){5}{4}
\Vertex(30,150){2}
\Text(18,120)[]{$r, \rho, k_1$}
\Text(0,65)[]{$t, \tau, k_3$}
\Text(46,65)[]{$s, \sigma, k_2$}
\Text(140,98)[]{$-ig f_{rst} V^{\rho \sigma \tau }\left(k_1,k_2,k_3 \right) $}
\ArrowLine(0,230)(40,260)
\ArrowLine(40,260)(80,230)
\Gluon(40,260)(40,300){5}{4}
\Vertex(40,260){2}
\Text(24,191.2)[]{$k,b, \nu$}
\Text(0,132)[]{$p,i, \gamma$}
\Text(48,132)[]{$q,j, \beta$}
\Text(135,162)[]{$ig t^b_{ji}\left( \gamma_{\nu} \right)_{\beta \gamma } $}
\end{picture} }\\
{\sl  Coupling of the fields in QCD}
\end{center}
The recursive content of the DS equation, for the gluon field for example,
can be understood diagrammatically as follows: 
\vspace*{0.5cm}
\begin{center}
\SetScale{0.6}
\fbox{
\begin{picture}(295,70)(0,0)
\Gluon(0,50)(30,50){5}{3}
\GCirc(40,50){10}{0.5}
\Text(12,42)[]{$P$}
\PText(60,47)(0)[l]{=}
\Gluon(70,50)(100,50){5}{3}
\Text(53,42)[]{$P$}
\PText(110,47)(0)[l]{+}
\Gluon(120,50)(150,50){5}{3}
\Gluon(150,50)(180,70){5}{3}
\Gluon(150,50)(180,30){5}{3}
\Vertex(150,50){2}
\GCirc(185,75){10}{0.5}
\GCirc(185,25){10}{0.5}
\Text(124,45)[]{$p_1$}
\Text(124,15)[]{$p_2$}
\PText(220,47)(0)[l]{+}
\Gluon(230,50)(270,50){5}{3}
\Gluon(270,50)(300,90){5}{4}
\Gluon(270,50)(310,50){5}{3}
\Gluon(270,50)(300,20){5}{4}
\Vertex(270,50){2}
\GCirc(310,95){10}{0.5}
\GCirc(320,50){10}{0.5}
\GCirc(305,15){10}{0.5}
\Text(200,56)[]{$p_1$}
\Text(206,30)[]{$p_2$}
\Text(200,8)[]{$p_3$}
\PText(370,47)(0)[l]{+}
\Gluon(380,50)(410,50){5}{3}
\ArrowLine(410,50)(430,65)
\ArrowLine(430,35)(410,50)
\Vertex(410,50){2}
\GCirc(435,70){10}{0.5}
\GCirc(438,28){10}{0.5}
\Text(280,42)[]{$p_1$}
\Text(280,15)[]{$p_2$}
\end{picture}   }\\
{\sl  Recursion equation for gluons}
\end{center}
The figure shows that a subamplitude with an off-shell gluon of momentum $P$,
has contributions from three- and four-vertices plus a fermion-antifermion vertex.
The shaded blobs denote subamplitudes with the same structure. So, in effect,
this is a recursive equation which we can write down immediately as (suppressing the
colour): 
\bqa
A^\mu(P)=\sum_{i=1}^n \delta_{P=p_i} A^\mu(p_i) &+& (ig_s) \sum_{P=p_1+p_2}\Pi^\mu_\nu
\bar{\psi}(p_1) \gamma^\nu \psi(p_2) \sigma(p_1,p_2) \nn \\
&+& \frac{(ig_s)}{2} \sum_{P=p_1+p_2} V^{\mu \nu \lambda}(P,p_1,p_2) A_\nu(p_1) A_\lambda(p_2) 
\sigma(p_1,p_2) \nn \\
&-& \frac{(g_s^2)}{6} \sum_{P=p_1+p_2+p_3} G^{\mu \nu \lambda \rho}
 A_\nu(p_1) A_\lambda(p_2) A_\rho(p_3) \sigma(p_1,p_2)
\label{Eq-glu}
\eqa
where $V^{\mu \nu \lambda}(P,p_1,p_2)$ and $G^{\mu \nu \lambda \rho}$ are the
3 and 4-point vertices and 
\[
\Pi^\mu_\nu=\frac{-i g^\mu_\nu}{P^2}  \nn 
\]
is the gluon propagator. The symbol $\sigma(p_1,p_2)$ 
- we will call it the sign function - takes into account the Fermi sign and has
the values $\pm1$. More on the sign function will be presented later on.
For a quark of momentum $P$ (suppressing the colour again)
\bq
\psi(P)=\sum_{i=1}^n \delta_{P=p_i} \psi(p_i)+(ig_s) \sum_{P=p_1+p_2}S 
 A\cfeyn (p_1) \psi(p_2) \sigma(p_1,p_2)
\label{Eq-qu}
\eq
\begin{center}
\SetScale{0.6}
\fbox{
\begin{picture}(140,60)(0,0)
\ArrowLine(0,50)(30,50)
\GCirc(40,50){10}{0.5}
\Text(12,38)[]{$P$}
\PText(60,47)(0)[l]{=}
\ArrowLine(70,50)(100,50)
\Text(53,38)[]{$P$}
\PText(110,47)(0)[l]{+}
\ArrowLine(120,50)(150,50)
\ArrowLine(150,50)(180,70)
\Vertex(150,50){2}
\Gluon(150,50)(180,30){5}{3}
\GCirc(185,75){10}{0.5}
\GCirc(185,25){10}{0.5}
\Text(124,47)[]{$p_1$}
\Text(124,15)[]{$p_2$}
\end{picture} 
}\\
{\sl  Recursion equation for quarks}
\end{center}   
\vspace*{-0.2cm}
where $S$ is the propagator
\[
S=\frac{iP \cfeyn}{ P^2} \nn
\]
and for an antiquark
\bq
\bar{\psi}(P)=\sum_{i=1}^n \delta_{P=p_i} \bar{\psi}(p_i)+(ig_s) \sum_{P=p_1+p_2} 
 \bar{\psi}(p_2) A\cfeyn (p_1) \tilde{S} \sigma(p_1,p_2)
\label{Eq-aqu}
\eq
\begin{center}
\SetScale{0.6}
\fbox{
\begin{picture}(140,60)(0,0)
\ArrowLine(30,50)(0,50)
\GCirc(40,50){10}{0.5}
\Text(12,38)[]{$P$}
\PText(60,47)(0)[l]{=}
\ArrowLine(100,50)(70,50)
\Text(52,38)[]{$P$}
\PText(110,47)(0)[l]{+}
\ArrowLine(150,50)(120,50)
\ArrowLine(180,70)(150,50)
\Vertex(150,50){2}
\Gluon(150,50)(180,30){5}{3}
\GCirc(185,75){10}{0.5}
\GCirc(185,25){10}{0.5}
\Text(124,45)[]{$p_1$}
\Text(124,10)[]{$p_2$}
\end{picture}   }\\
{\sl   Recursion equation for antiquarks}
\end{center}
\vspace*{-0.3cm}
where
\[
\tilde{S}=\frac{-iP \cfeyn}{ P^2 }  \nn
\]
In order to reduce computational complexity, as we will discuss later on, we
replace the four gluon vertex with a three-vertex by introducing an auxiliary field
$H_{\mu \nu}$. We rewrite the part of the QCD Lagrangian that describes the four-vertex
in terms of the auxiliary field as follows:
\bq 
{\cal L}_H= - g f^{abc} A_{ \mu}^{a} A_{ \nu}^{b} H^{ \mu \nu \: c}
- H_{ \mu \nu }^{a}  H^{ \mu \nu \: a } 
\eq
The recursion for the gluons changes slightly, namely only the part with the four-vertex.
Additionally, we also have an equation for the auxiliary field:
\begin{center}
\SetScale{0.6}
\fbox{
\begin{picture}(190,140)(0,0)
\DashLine(65,50)(95,50){3}
\GCirc(105,50){10}{0.5}
\Text(50,37)[]{$P$}
\PText(125,45)(0)[l]{=}
\DashLine(135,50)(165,50){3}
\Gluon(165,50)(195,70){5}{3}
\Gluon(165,50)(195,30){5}{3}
\GCirc(200,75){10}{0.5}
\GCirc(200,25){10}{0.5}
\Vertex(165,50){2}
\Text(135,46)[]{$p_1$}
\Text(135,15)[]{$p_2$}
\Gluon(0,150)(30,150){5}{3}
\Gluon(30,150)(60,190){5}{4}
\Gluon(30,150)(70,150){5}{3}
\Gluon(30,150)(60,120){5}{4}
\Vertex(30,150){2}
\GCirc(70,195){10}{0.5}
\GCirc(80,150){10}{0.5}
\GCirc(65,115){10}{0.5}
\Text(57,116)[]{$p_1$}
\Text(63,90)[]{$p_2$}
\Text(52,68)[]{$p_3$}
\PText(125,150)(0)[l]{=}
\Gluon(140,150)(170,150){5}{3}
\Gluon(170,150)(200,190){5}{4}
\DashLine(170,150)(210,150){3}
\Gluon(210,150)(250,150){5}{4}
\Gluon(210,150)(240,120){5}{4}
\Vertex(170,150){2}
\Vertex(210,150){2}
\GCirc(210,195){10}{0.5}
\GCirc(260,150){10}{0.5}
\GCirc(245,115){10}{0.5}
\Text(143,115)[]{$p_1$}
\Text(170,90)[]{$p_2$}
\Text(160,67)[]{$p_3$}
\Text(115,80)[]{$H$}
\end{picture}  }\\
{\sl   Elimination of the four-vertex and the new H-gluon-gluon vertex.}  
\end{center}
\bqa
A^\mu(P)=\sum_{i=1}^n \delta_{P=p_i} A^\mu(p_i) &+& (ig_s) \sum_{P=p_1+p_2}\Pi^\mu_\nu
\bar{\psi}(p_1) \gamma^\nu \psi(p_2) \sigma(p_1,p_2) \nn \\
&+& \frac{(ig_s)}{2} \sum_{P=p_1+p_2} V^{\mu \nu \lambda}(P,p_1,p_2) A_\nu(p_1) A_\lambda(p_2) 
\sigma(p_1,p_2) \nn \\
&-& {(g_s)} \sum_{P=p_1+p_2} X^{\mu \nu \lambda \rho}
 A_\nu(p_1) H_{\lambda \rho}(p_2)  \sigma(p_1,p_2) \nn \\
\label{Eq-glu1}
\eqa
\bq
H_{\mu \nu}(P)=- \frac{(g_s)}{4} \sum_{P=p_1+p_2} X^{\mu \nu \lambda \rho}
A_\lambda(p_1) A_\rho(p_2) \sigma(p_1,p_2)
\label{Eq-aux}
\eq
where $X^{\mu \nu \lambda \rho}$ is the new H-gluon-gluon vertex:
\[
X^{\mu \nu \lambda \rho}=g^{\mu \lambda}g^{\nu \rho}-g^{\nu \lambda}g^{\mu \rho} \nn
\]
These four equations, namely (\ref{Eq-glu1}),(\ref{Eq-aux}),(\ref{Eq-qu}),(\ref{Eq-aqu}),
represent off-shell subamplitudes
that are the building blocks of any process. They are used iteratively, combining
two (or three) momenta, at each step, to build a subamplitude.
The iteration begins with the initial conditions for
the external particles. In particular for a gluon we have:
\bq 
A_a^\mu(p_i) = \epsilon^\mu_\lambda(p_i) \, \delta_{aa_i} \; , \; \; \; \; i=1,..n
\eq 
where $a$ is the colour, $a=1,\ldots,8$, $\epsilon^\mu_\lambda(p)$ denotes the polarization 
vector and $i$ denotes one the external gluons. For the 
quarks and antiquarks the iteration starts with
\bqa
\psi_{k}(p_i) =  \left \{ \begin{array}{ll} 
                             u(p_i) \delta_{kk_i} 
                             &  \textrm{if} \; i \; \textrm{incoming} \\[12pt] 
                             \bar{u}(p_i) \delta_{kk_i}  
                             & \textrm{if} \; i \; \textrm{outgoing}
                                           \end{array} 
                                        \right. \nn \\
\bar{\psi}_{k}(p_i) =  \left \{ \begin{array}{ll} 
                             \bar{u}(p_i) \delta_{kk_i} 
                             &  \textrm{if} \; i \; \textrm{incoming} \\[12pt] 
                             u(p_i) \delta_{kk_i}  
                             & \textrm{if} \; i \; \textrm{outgoing}
                                           \end{array} 
                                        \right. 
\eqa
where $k$ is the colour, $k=1,2,3$.
The next step is combining two of the external momenta, in all possible ways, in accordance
to the Feynman rules, to compute the next subamplitude. 
The iteration goes through in the same manner, using the equations (\ref{Eq-glu1}),
(\ref{Eq-aux}),(\ref{Eq-qu}),(\ref{Eq-aqu}) repeatedly, each time combining the new 
momenta obtained,
with the remaining of the external ones. After $n-1$ steps there is only one momentum
left to be combined, so the last step gives us the amplitude 
$\mathcal{A}(p_1,p_2, \ldots ,p_n)$. We refer to reference~\cite{helac} for all further details
of the algorithm.
\subsection{Colour and helicity}
In order to have an estimate of the production probability, one has to
sum over all colour and helicity configurations.
Summation over colours is a delicate subject. If one performs the summation 
in a straightforward way then one has to consider something like
$8^{n_g} \times 3^{n_q} \times 3^{n_{\bar q} }$ configurations for the $n$-parton scattering,
where $n_g,n_q,n_{\bar q}$ is the number of gluons, quarks and antiquarks respectively. 
In this section we show how this summation can be replaced by 
integration, which is
then suitable for Monte Carlo computation.
As a first step a simplification of the colour structure is possible by defining 
the following object~\cite{DragPap}
\bq
G_{AB}\equiv \sum_{a=1}^{8} t^a_{AB} G^a, \;\;\; A,B=1,2,3
\label{Color}
\eq
where $G^a$ is the gluon field and all other indices have been temporarily suppressed. 
The new objects are of course traceless, $3 \times 3$ matrices in colour space. The 
interesting property of this
colour representation is that it leads to a ``diagonalization''
of the colour structure of the three-gluon vertex. More specifically the
colour part of the three-gluon vertex is now given by
\bq
f^{abc}t^a_{AB}t^b_{CD}t^c_{EF}=
- \frac{i}{4} (\delta_{A D} \: \delta_{C F} \: \delta_{E B} 
- \delta_{A F} \: \delta_{C B} \: \delta_{E D}) 
\label{diag-col}
\eq
\begin{center}
\fbox{
\begin{picture}(270,95)(0,0)
\ArrowArc(20,20)(40,0,80)
\ArrowArc(107,20)(40,95,180)
\ArrowArc(65,105)(50,225,315)
\Text(20,60)[]{${\it B }$}
\Text(25,75)[]{${\it A }$}
\Text(57,14)[]{${\it E }$}
\Text(70,14)[]{${\it F }$}
\Text(105,75)[]{${\it D }$}
\Text(110,60)[]{${\it C }$}
\PText(130,40)(0)[l]{___}
\ArrowArcn(150,20)(40,80,360)
\ArrowArcn(237,20)(40,180,95)
\ArrowArcn(195,105)(50,315,225)
\Text(150,60)[]{${\it A }$}
\Text(155,75)[]{${\it B }$}
\Text(187,14)[]{${\it F }$}
\Text(200,14)[]{${\it E }$}
\Text(235,75)[]{${\it C }$}
\Text(240,60)[]{${\it D }$}
\end{picture}
}\\
{\sl  Colour flow in the gluon 3-vertex, as represented in Eq.(\ref{diag-col}).}
\end{center}
This colour structure shows the colour flow in the real physical
process, where gluons can be represented by quark-antiquark states in colour
space and their self-interaction, as given by \eqn{diag-col}, reflects 
the fact that
colour remains unchanged on an uninterrupted colour line. The recursion equations that
include the gluon, like Eq.(\ref{Eq-glu1}), now are modified according to Eq.(\ref{Color}),
to reflect the new colour structure. The full content
of the recursion equations, including the colour structure as just described,
is listed in the Appendix A.

This new, simplified colour structure of the vertices, allows us now to
take the next step in making the computation of the colour part of an amplitude
more efficient, by ridding ourselves of the summations, mentioned in the
beginning of this section, and replacing them by integration. To this end,
we assign to a fermion a complex vector $z_A$, where the index runs from 1 to 3,
representing its colour content. These vectors parametrize the 5-dimensional 
representation of $SU(3)$ on the sphere and are subjected to the constraint: 
\[ 
\mid z_1 \mid^{2}+\mid z_2 \mid^2+\mid z_3 \mid^2 =1 \nn
\] 
In this space, integration is defined through the proper definition for the
invariant group measure, $[dz]$:
\[
\int [dz] \equiv \int \left( \prod_{i=1}^{3}dz_i dz_i^*\right)  
\delta( \sum_{i=1}^{3} z_i z_i^* - 1)  \nn
\]
We can use a simple polar coordinates parametrization to represent these
complex vectors:
\bqa 
&z_1=e^{i \phi_1} \cos \theta & \nn \\ 
&z_2=e^{i \phi_2} \sin \theta \cos \xi &  \nn \\ 
&z_3=e^{i \phi_3} \sin \theta \sin \xi &   \nn \\
&0 \leq \phi_i \leq 2 \pi  \; \; ; \; \; 0 \leq \theta \leq \frac{ \pi }{2}  
 \; \; ; \; \;    0 \leq \xi \leq  
\frac{ \pi }{2} &
\eqa 
and in terms of these variables, the invariant measure becomes:
\[ 
\frac{1}{\pi^3}
\left( \prod_{i=1}^{3} \int_0^{2 \pi} d\phi_i \right)
\int_0^{ \frac{ \pi }{2} } d \theta 
\int_0^{ \frac{ \pi }{2} } d \xi 
\cos \theta \sin^3 \theta \cos \xi \sin \xi   \nn
\] 

The colour structure as described by (\ref{diag-col}), shows that gluons can be interpreted as 
quark-antiquark pairs. So we construct the following vector, appropriate
for describing the colour part of a gluon, which has this colour structure:
\bq 
\eta^{a} (z)= \sqrt{24} \sum_{i,j=1}^{3} z_i^{*} ( t^{a} )_{ij} z_j   
\; \; \; \; \; a=1,\ldots,8 
\eq
where $t^a$ are the Gell-Mann matrices. 
This vector is real, because $(\eta^{a})^{\ast}=\eta^{a}$ due to the 
hermiticity of the Gell-Mann matrices, and is normalized as follows:
\[
\int [dz]\, \eta^a(z)\eta^b(z) = \delta^{ab}
\label{eta-norm}
\]
  
As far as our recursive equations are concerned their structure remains 
unaffected
and the only thing to be changed are the initial conditions :
\bqa
&&G_{AB}^{\mu}(P_i)=\sum_{a=1}^{8}G^a(P_i) \eta^a(z)
=  \sqrt{6} \left(
z_{iA} z^*_{iB}-\frac{1}{3} \delta_{AB} \right) \epsilon^\mu_\lambda(P_i) \nn \\
&&\psi_A(P_i)= \sqrt{3} \; u(P_i) \; z_{iA} \nn \\
&&\bar{\psi}_A(P_i)= \sqrt{3} \; \bar{u}(P_i) \; z^*_{iA}
\eqa
where as usual $i=1,\ldots,n$, $\lambda$ is the helicity
and $\vec{z}_i$ are the new continuous colour 
coordinates of the $i$-th parton. The constants in the front are normalizations.

In the same spirit summation over helicity configurations of the external
partons can be replaced by an integration over a phase variable. For example, 
for a gluon this
is achieved by introducing the polarization vector
\[
\epsilon_\phi^\mu(p)=
e^{i\phi} \epsilon^\mu(p,+)+e^{-i\phi} \epsilon^\mu(p,-) \nn
\]
where $\phi$ is a random number.
Then by integrating over $\phi$ we obtain the sum over helicities,
\[
\frac{1}{\pi}\int_{0}^{\pi}d\phi\,\epsilon_\phi^\mu(p) 
(\epsilon_\phi^\nu(p))^*=
\sum_{\lambda=\pm} \epsilon^\mu(p,\lambda) (\epsilon^\nu(p,\lambda))^*  \nn
\]
The same thing can be used for the helicity of quark or an anti-quark. For example, 
for the quark we have
\[
u_{\phi}(p)=e^{i\phi} u_{+}(p) + e^{-i\phi} u_{-}(p) \nn
\]
and when integrated over $\phi$ it gives the sum over polarizations
\[
\sum_{\lambda=\pm} u_{\lambda}(p) \bar{u}_{\lambda}(p)=p\feyn \nn
\]
and the same for an antiquark.
\subsection{The Fermi sign function}
Since we are dealing with fermions, we must find a way to incorporate 
a sign change when we interchange
two identical fermions in a process. To this end we use the binary representation
of the momentum labels of the external particles (e.g. $P_1 \rightarrow (0001),
P_2 \rightarrow (0010), P_3=P_1+P_2 \rightarrow (0011)$, etc.) So the sign relative
to the permutation of two momenta, $\sigma(P_i,P_j)$ is computed as an operation
on the two binary strings representing those momenta, $\sigma(m_i,m_j)$. The 
function that performs that operation is defined as:
\bq
\sigma(m_1,m_2)=(-1)^{\chi \left(m_1,m_2\right)}
\eq
with
\bq
\chi \left(m_1,m_2\right)=\sum_{i=n}^{2} \hat{m}_{1i} 
\left( \sum_{j=1}^{i-1} \hat{m}_{2j} \right)
\eq
A hat over the binary string means that this particular bit is set to $0$ 
if the corresponding external particle is a boson.

\subsection{Flavor treatment}
The classification of processes contributing to the production of $n$ jets is simplified
when certain symmetries are taken into account. For instance processes like $gg\to u\bar{u}
u \bar{u}$ and $gg\to d\bar{d}d\bar{d}$ can be grouped together in the general class with
representative 
$gg\to q\bar{q}q\bar{q}$; obviously an extra factor $f$, representing the number of light flavours,
has to be taken into account. We may call the  representative `a distinct processes' and attempt
a complete classification of such processes. Is clear that the number of  distinct processes for 
a given number $n$ of jets will be a small fraction of the total number of processes; this 
has a major implication on the computational complexity of the problem, since we may have
the total contribution by just computing the contribution of a relatively small number of distinct
processes.
In the following table we give the number of distinct processes, as defined in Appendix C,
for $f=5$ final state flavours, 
and $f=4$ initial state flavours, where we used these numbers in order to be able to 
compare with ref.~\cite{Kuijf}.

\begin{center}
\begin{tabular}{|c|c|c|c|c|c|c|c|c|c|}
\hline
\# of jets & 2 & 3 & 4 & 5 & 6 & 7 & 8 & 9 & 10
\\
\hline
\# of dist.processes  & 
10  & 14   & 28  & 36 & 64    & 78    & 130   & 154  & 241
\\
\hline
total \# of processes  &
126 & 206  & 621 & 861 & 1862 & 2326  & 4342  & 5142 & 8641
\\
\hline
\end{tabular} 
\end{center}

Another way to tackle all these different processes is
to introduce quarks (antiquarks) that are a mixture of different flavours. 
Thus, keeping with the spirit of colour and helicity treatment,
we integrate over flavours instead of summing over them. This is done by attaching
a vector $\vec{f}$ to each spinor describing a fermion, the components
of which are random numbers between 0 and 1:
\[
\psi_p=u(p) \times \vec{f}  \nn
\]  
with
\[
\vec{f}=\sqrt{N_f} \; \left( \begin{array}{c} f_1\\f_2\\f_3\\ \vdots \\f_{N_f} \end{array} \right) \nn
\]
where $f_1,f_2,\ldots,f_{N_f}$ are random numbers and 
$N_f$ is the number of flavours. We choose the vectors $\vec{f}$ in such
a way so that they are normalized as follows:
\[
\langle f_i f_j \rangle = \delta_{ij} \nn 
\]
where $i,j=1,\ldots,N_f$. As far as the final state is concerned, all flavours
are equally treated (massless quarks). In the initial state however, due to different
structure functions, special care should be taken of different flavours 
by a suitable weighting of the initial conditions with the structure
function appropriate for each flavour.

In treating flavour as described so far a new `definition' of the concept of
distinct process emerges. Any process is now composed by three primary objects, namely
$g$, $q$ and $\bar{q}$,  where $q$($\bar{q}$) 
represents a coherent superposition of all flavour states. The possible initial states are 
nine, namely 
$gg,gq,g\bar{q},qg,\bar{q}g,q\bar{q},\bar{q}q,qq,\bar{q}\bar{q}$
and a final state will be determined by fixing the number $m$ of $q\bar{q}$ pairs.
Obviously $m$ should satisfy,  $2m \leq n-c_i$, with $c_i$ counts the net quark (antiquark)
content of the initial state, namely 
\[
c_{gg}=c_{q\bar{q}}=c_{\bar{q}q}=0\;\;\;c_{gq}=c_{qg}=c_{\bar{q}g}=c_{g\bar{q}}=1\;\;\;
c_{qq}=c_{\bar{q}\bar{q}}=2 
\]
So the number of 'distinct processes' is now given by
\bq 
9 k + 3\;\; \mbox{if} \;\;n=2 k\;\;\;\mbox{and}\;\;\;
 9 k + 7 \;\; \mbox{if}\;\; n=2k +1 \nn
\label{distproc}
\eq
which represents a further reduction compared with 
the number of distinct 
processes indicated in the previous table.
\section{Jet production rates}

The features discussed in the previous sections have been implemented
in a {\tt FORTRAN} code with which we are able to perform perturbative
calculations in QCD. We can divide the structure of the calculation leading to 
jet production rates  in four parts: event generation, phase-space generation,
squared amplitude computation and sum over processes calculation. In somewhat more
detail these parts consist of:
\paragraph{Amplitude computation} The basic structure of the algorithm for the
amplitude was discussed in section 2. In order to obtain squared amplitudes
we use projection to a continuous colour space and colour MC integration and
integration over helicity states. What remains to be done is the ability
to extract information about the colour string of the final states
so that it can be used in codes like {\tt HERWIG} \cite{herwig}. Work is being done
towards this end \cite{colourstrings}.
\paragraph{Phase space generation} In the current status of the code we 
have implemented two ways of phase-space generation. First of all we
have {\tt RAMBO} \cite{rambo}, a flat phase space generator which has been around
for many years and has proved reliable for the relevant calculations. Recently
though, a new generator has been developed, {\tt SARGE} \cite{HaKl}, which take 
into account the so called antenna structure of QCD amplitudes.
So far it has been proved very efficient for the generation of this type of
phase-space  \cite{HKD}.
\paragraph{Event generation} Let us consider the scattering of two hadrons.
The content of a hadron is characterized by the parton structure
function $f(x,Q^2)$ where $x$ is the fraction of the momentum $P$ of the hadron,
carried by the parton, $p=xP$, and $Q^2$ is the QCD scale.
The cross section for the scattering of two hadrons is given by the sum of all subprocesses
between the parton constituents of the hadrons, weighted with the corresponding
structure functions of the incoming partons:  
\bq
\sigma(s)=\sum_{ij\,P}  \int_{0}^{1} dx_1 dx_2 \int d\Phi \;  F_i(x_1) \;  F_j(x_2) \;
\left( \frac{ d \hat{\sigma} }{ d\Phi} \right)_{ij}
\label{Total-cross}
\eq
where $s$ is the centre of mass(CM) energy squared at the hadron level.  
As usual the functions $F_m(x)$ are defined as $F_m(x)=f_m(x)/x$, where $f_m(x)$ are the various
parton structure functions. The $\left( \frac{ d \hat{\sigma} }{ d\Phi} \right)_{ij}$ 
is the matrix element squared, summed (i.e. integrated) over helicity 
and colour degrees of freedom, and 
$d\Phi$ is the element of the phase space. The sum (i.e. integral) is over all partonic processes. 
We use Eq.(\ref{Total-cross}) to estimate jet production rates.

\paragraph{Sum over processes} Finally, once given the number of jets, 
we are able to compute all the relevant subprocesses that contribute at one go.
This is done by randomly choosing a subprocess
and then using Monte-Carlo to obtain the total cross section from all contributions.
The random choice of a subprocess is based on the choice of a pair of integers
$(i,m)$, $i$ selecting one out of the nine possible initial states,
and $m$ being the number of $q\bar{q}$ pairs in the final state, see \eqn{distproc}.

There is  also the option to use an optimization based on the fact that 
some processes overwhelm the total
cross section over others (for example the purely gluonic process,
$gg \to ng$ has the largest cross section by an order of magnitude, compared
to processes with different initial or final states).  

The algorithm has been used firstly to compute cross sections for 4 and 5
jet production.
We have chosen a CM energy of $\sqrt{s}=14$ TeV and 
we apply the following cuts:
\[
p_{T \; i} > 60 \; GeV, \; \; \; \; \; \theta_{ij} > 30^{o} \; \; \; \; \; |\eta_i| < 3 \nn
\label{cuts}
\]
where $p_T=\sqrt{p_x^2+p_y^2}$ is the transverse momentum of a jet, 
$\theta_{ij}=\measuredangle \left(\vec{p}_i,\vec{p}_j \right)$, is the
angle between jets and $\eta= -\ln \tan \left(\frac{\theta}{2} \right)$, is  the
pseudo-rapidity. 

For convenience all results are obtained with a non-running strong coupling constant put equal
to unity.
There are several parametrizations for the parton structure functions in the literature.
The one that we will use is the MRST parametrization \cite{MRST}, and the number of light flavours
is taken to be $f=5$.

In Table 1 we have listed several subprocesses and their cross sections, relative
for $pp$ scattering at LHC. The table is organized as follows:
we are not referring to a particular quark flavour, so we use the letters $q,r,s$ to denote
quarks or antiquarks. Since we consider all quarks and antiquarks massless it
makes no difference which particular flavour appears in the final state. 
We compare our results with those from a well established code, {\tt NJETS}.
We also show distributions of the maximum $p_T$ for two processes, namely $gg \to gggg$
and $q\bar{q} \to r\bar{r} s\bar{s}$. 

To show the potential of the flavour integration method we
calculate the total cross section for a set of  possible final states
and compare with {\tt NJETS} where as usual a sum over different 
flavour states has been used. The results are presented in Table 2. 
We have used the letter $q$ to
denote any flavour and all combinations have been taken into account in the calculation
of the cross section. We have used the same cuts and CM energy as before.

We have also obtained the total cross sections for the production of 3,4,5,6,7 and 8 
jets. These are listed in Table 3. Also shown there is the contribution of the
purely gluonic process to the total cross section. We see the fraction of this process
compared to the total is diminishing with increasing number of jets, as was already
commented in \cite{DK}.

Finally we have plotted distributions of the total cross section for the production 
of 5,6 and 7 jets. The quantities shown are:
\begin{itemize}
 \item Transverse momenta $p_T$ of the products
 \item Invariant masses  $M_{ij}=\sqrt{2 (p_i \cdot p_j) }$ 
\end{itemize}

\section{Summary}
We have presented a procedure to calculate matrix elements and cross sections
in QCD using an iterative algorithm based on the Schwinger-Dyson equation. 
Thus we free ourselves from the task of
computing all Feynman graphs for a process, a task which can become impossible even
for a moderate number of particles involved. We have also managed to bypass the 
computationally expensive procedure of summing over all possible colour and helicity
configurations, by introducing continuous vectors to represent these otherwise discrete
quantities and using Monte-Carlo integration instead of summation. At this stage our code
can reliably compute scattering amplitudes and cross sections
\begin{itemize}
 \item for individual, single-flavour processes 
 \item for processes where a sum over all contributing flavours is needed
       both in the initial or the final states
 \item for total jet cross sections, where we are interested in production rates
       with contribution from all possible subprocesses and all flavours.
\end{itemize}
Our future interests involve a convolution of this code with fragmentation 
codes like {\tt HERWIG}, so that one can perform realistic simulations 
of multi-jet processes.

\newpage

\begin{center}
\begin{tabular}{|>{$}l<{$}||>{$}c<{$}|>{$}c<{$}|}
  \hline
  \multicolumn{3}{|>{$}c<{$}|}{\sqrt{s}= 14 \; \mathrm{TeV}} \\
  \hline  
  \mathrm{Process} &{\tt OUR \; CODE} \mathrm{(nb)} &{\tt NJETS}\mathrm{(nb)}  \\
\hline   
\hline   
  gg \to gggg &2.681  &2.533   \\
  q\bar{q} \to gggg &0.0020  &0.0021   \\
  qg \to qggg &1.131  &1.159    \\
  gg \to q\bar{q} gg &0.106 &0.104   \\ 
  q\bar{q} \to q\bar{q} gg &0.062 &0.059   \\ 
  q\bar{q} \to r\bar{r} gg &0.586 \times 10^{-3} &0.558 \times 10^{-3}  \\ 
  qq \to qq  gg &0.134 &0.126   \\ 
  q\bar{r} \to q\bar{r}  gg &0.171 &0.161   \\ 
  qr \to qr  gg &0.210 & 0.197  \\ 
  qg \to q\bar{q}q g &0.035 &0.033   \\ 
  qg \to r\bar{r}q g &0.032 &0.035  \\ 
  gg \to q\bar{q}q\bar{q}  & 0.524 \times 10^{-3} &0.526 \times 10^{-3}   \\ 
  gg \to q\bar{q}r\bar{r}  &1.059 \times 10^{-3}&1.074 \times 10^{-3}   \\ 
  q\bar{q} \to q\bar{q}q\bar{q}  &0.807 \times 10^{-3} &0.851 \times 10^{-3}   \\ 
  q\bar{q} \to q\bar{q}r\bar{r}  &0.866 \times 10^{-3} &0.786 \times 10^{-3}   \\ 
  q\bar{q} \to r\bar{r}r\bar{r}  & 7.94 \times 10^{-6} &7.93 \times 10^{-6}   \\ 
  q\bar{q} \to r\bar{r}s\bar{s}  & 1.54 \times 10^{-5} &1.49 \times 10^{-5}   \\ 
  qq \to qq q\bar{q}  &1.64 \times 10^{-3}  & 1.60 \times 10^{-3}  \\ 
  qq \to qq r\bar{r}  &1.67 \times 10^{-3} &1.60 \times 10^{-3}   \\ 
  qr \to qr q\bar{q}  &2.78 \times 10^{-3} &2.91 \times 10^{-3}  \\ 
  qr \to qr s\bar{s}  &2.63 \times 10^{-3} &2.57 \times 10^{-3}   \\ 
  q\bar{r} \to q\bar{r}q\bar{q}  &2.28 \times 10^{-3} &2.41 \times 10^{-3}   \\ 
  q\bar{r} \to q\bar{r}r\bar{r}  &2.20 \times 10^{-3} &2.07 \times 10^{-3}   \\ 
  q\bar{r} \to q\bar{r}s\bar{s}  &2.39 \times 10^{-3}&2.12 \times 10^{-3}   \\ 
\hline
\end{tabular}
\begin{tabular}{|>{$}l<{$}||>{$}c<{$}|>{$}c<{$}|}
  \hline
  \multicolumn{3}{|>{$}c<{$}|}{\sqrt{s}= 14 \; \mathrm{TeV}} \\
  \hline  
  \mathrm{Process} &{\tt OUR \; CODE} \mathrm{(pb)} &{\tt NJETS}\mathrm{(pb)}  \\
\hline   
\hline   
  gg \to ggggg &159.49  &160.40    \\
  q\bar{q} \to ggggg &0.141  & 0.136  \\
  qg \to qgggg &87.94  & 84.37   \\
  gg \to q\bar{q} ggg &9.193 &9.218   \\ 
  q\bar{q} \to q\bar{q} ggg &4.37 &4.561   \\ 
  q\bar{q} \to r\bar{r} ggg &0.034 &0.0348   \\ 
  qq \to qq  ggg &11.47 & 12.35  \\ 
  q\bar{r} \to q\bar{r}  ggg &11.39 & 11.98  \\ 
  qr \to qr  ggg &16.51 &17.29   \\ 
  qg \to q\bar{q}q gg &3.858 &3.799   \\ 
  qg \to r\bar{r}q gg &3.738 &3.887   \\ 
  gg \to q\bar{q}q\bar{q} g  &0.104 &0.101  \\ 
  gg \to q\bar{q}r\bar{r} g  &0.207  &0.206  \\ 
  q\bar{q} \to q\bar{q}q\bar{q} g  &0.252 & 0.259  \\ 
  q\bar{q} \to q\bar{q}r\bar{r} g  &0.230 &0.254  \\ 
  q\bar{q} \to r\bar{r}r\bar{r} g &0.0020 &0.0020    \\ 
  q\bar{q} \to r\bar{r}s\bar{s} g &0.0038 &0.0038    \\ 
  qq \to qq q\bar{q} g  & 0.684&0.691   \\ 
  qq \to qq r\bar{r} g &0.698 & 0.659 \\ 
  qr \to qr q\bar{q} g &0.998 &0.922   \\ 
  qr \to qr s\bar{s} g &0.862 &0.941   \\ 
  q\bar{r} \to q\bar{r}q\bar{q} g & 0.684 &0.658   \\ 
  q\bar{r} \to q\bar{r}r\bar{r} g &0.650 & 0.678  \\ 
  q\bar{r} \to q\bar{r}s\bar{s} g &0.634 &0.665   \\ 
  qg \to q\bar{q}q\bar{q} q &0.0352 &0.0334   \\ 
  qg \to q\bar{q}r\bar{r} q &0.0646 &0.0682   \\ 
  qg \to r\bar{r}r\bar{r} q &0.0328 &0.0328   \\ 
  qg \to r\bar{r}s\bar{s} q &0.0650 &0.0668   \\ 
\hline
\end{tabular}
\end{center}
\vspace*{-0.1cm}
\hspace*{1.8cm} {\bf Table 1} Production rates for 4 and 5 jet production. 
                              All results have
\newline 
\hspace*{1.8cm} an estimated error of  $4 \% $ .
\vspace*{0.5cm}

\newpage

\begin{center}
\begin{tabular}{|>{$}c<{$}||>{$}c<{$}|>{$}c<{$}|}
  \hline \multicolumn{3}{|>{$}c<{$}|}{\sqrt{s}= 14 \; \mathrm{TeV}  } \\
  \hline \mathrm{Process} &  {\tt OUR \; CODE}\; \mathrm{(nb)} & {\tt NJETS} \; \mathrm{(nb)}  \\
  \hline q\bar{q} \to q\bar{q} gg &0.242 & 0.232 \\
         q\bar{q} \to q\bar{q}q\bar{q} &0.015 &0.014  \\ 
         qq \to qq gg &0.311 &0.322  \\
         gg \to q\bar{q} q\bar{q} &0.013 & 0.014\\ 
         qg \to q\bar{q} qg&0.167 &0.174  
\\\hline
\end{tabular}
\end{center}
\hspace*{2.8cm} {\bf Table 2} Flavour integration with an estimated error of $2 \%$. 

\begin{center}
\begin{tabular}{|l|c|c|c|c|c|c|}\hline
$\# \; \mathrm{of} \; \mathrm{jets}$ & 3 & 4 & 5 & 6 & 7 & 8  \\
\hline \hline
$\sigma (nb)$  &91.41 &6.54 &0.458 &2.97 $\times 10^{-2}$ &2.21 $\times 10^{-3}$ &2.12 $\times 10^{-4}$  \\
\hline
$\% \; \mathrm{Gluonic}$ &45.7 &39.2 &35.7 &35.1 &33.8 &26.6  \\
\hline
\end{tabular}
\end{center}
\hspace*{2.8cm} {\bf Table 3} Total cross sections for the production of up to $8$ jets.
\newline
\hspace*{2.8cm}  The last row shows the percentage contribution of the purely
\newline
\hspace*{2.8cm}  gluonic process ( $gg \to ng$ ). The estimated error is $3 \%$ for $3$   
\newline
\hspace*{2.8cm}  and $4$ jets, $4 \%$ for $5$ and $6$ jets  and  $6 \%$ for $7$ and $8$ jets.  

\begin{figure}
\begin{center}
\epsfig{file=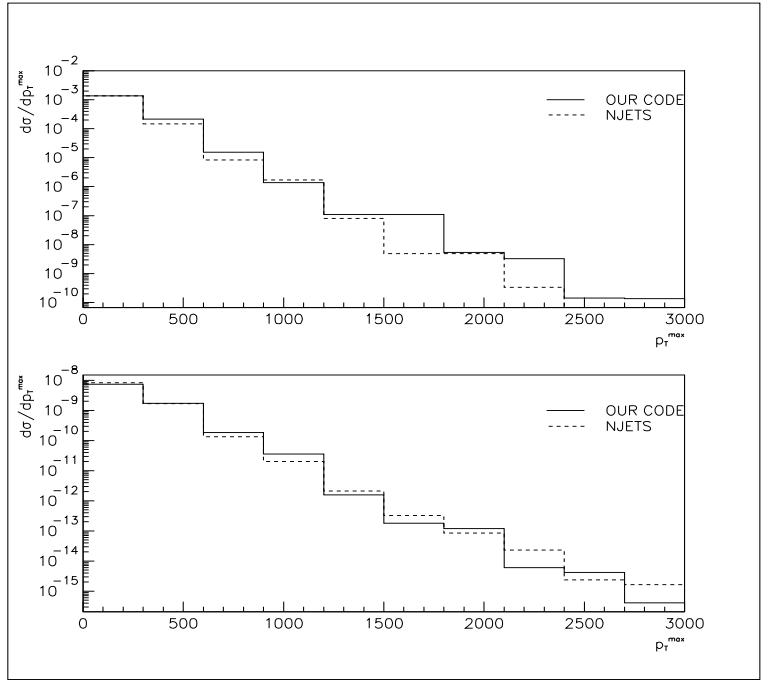,height=9cm,width=10cm}
\end{center}
\caption{Maximum $p_T$ distributions for $gg \to gggg$ (top plot) and
          $q\bar{q} \to r\bar{r} s\bar{s}$ (bottom plot)}
\end{figure}

\begin{figure}
\begin{center}
\epsfig{file=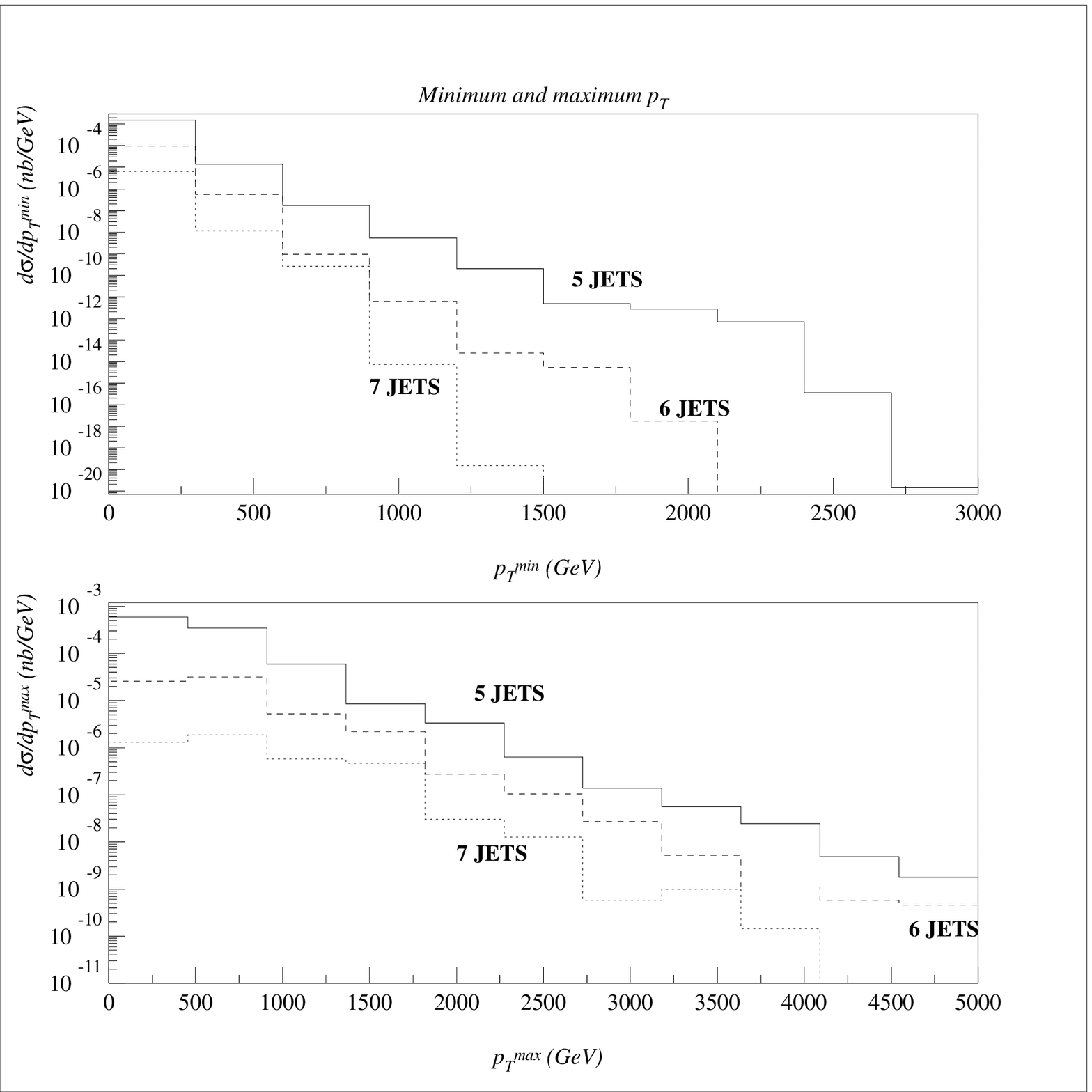,height=9cm,width=10cm}
\epsfig{file=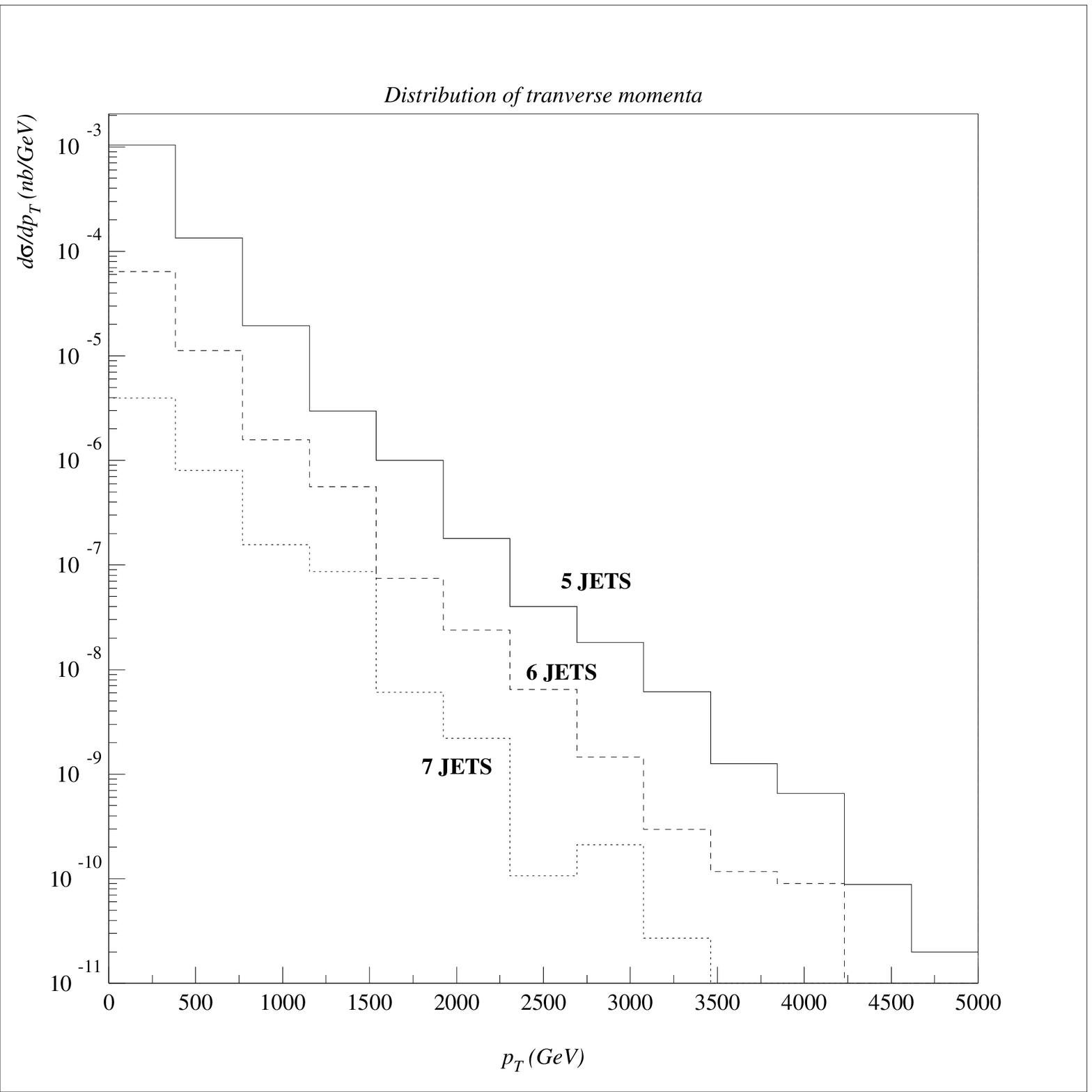,height=9cm,width=10cm}
\end{center}
\caption{$p_T$ distributions for 5,6 and 7 jet production}
\end{figure}

\begin{figure}
\begin{center}
\epsfig{file=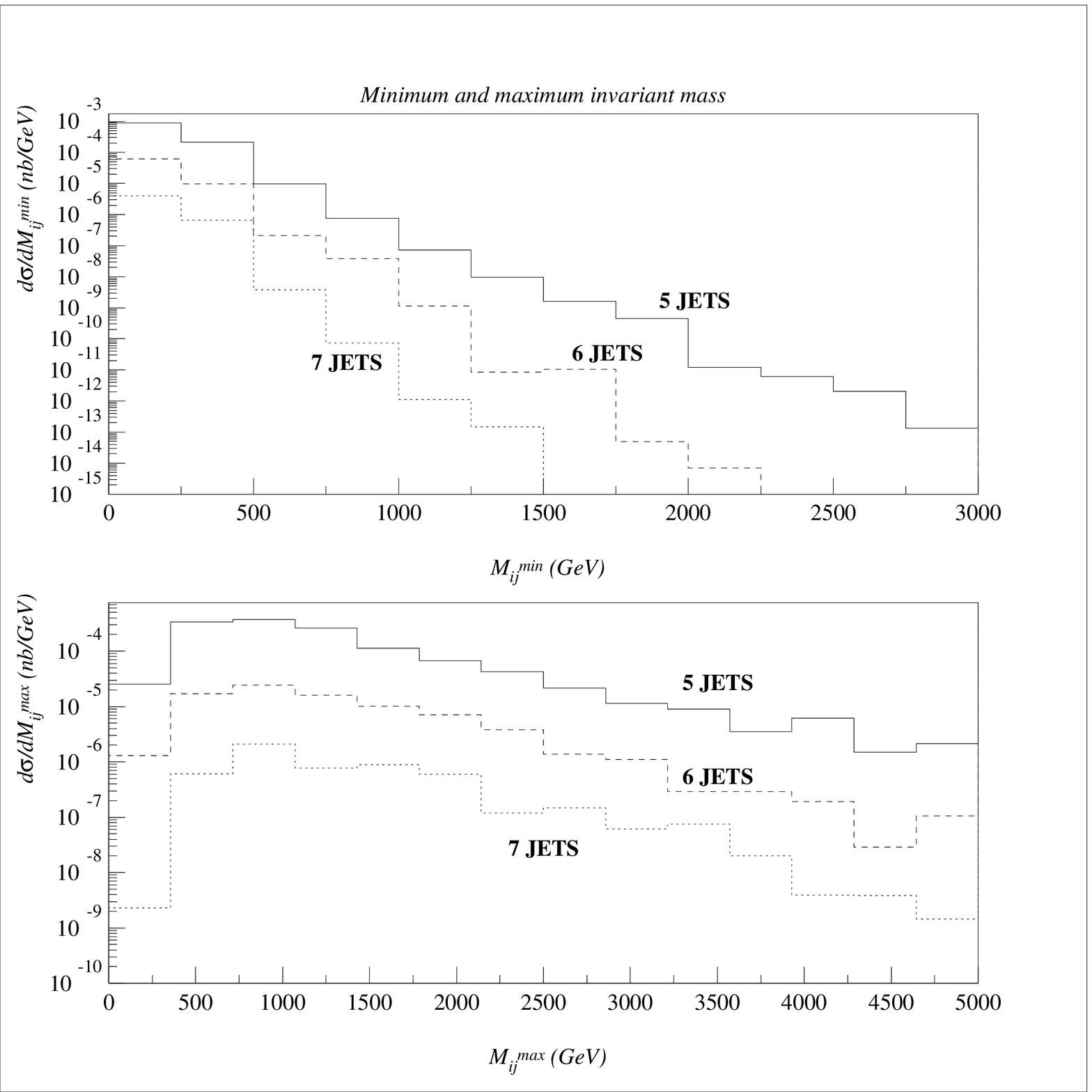,height=9cm,width=10cm}
\epsfig{file=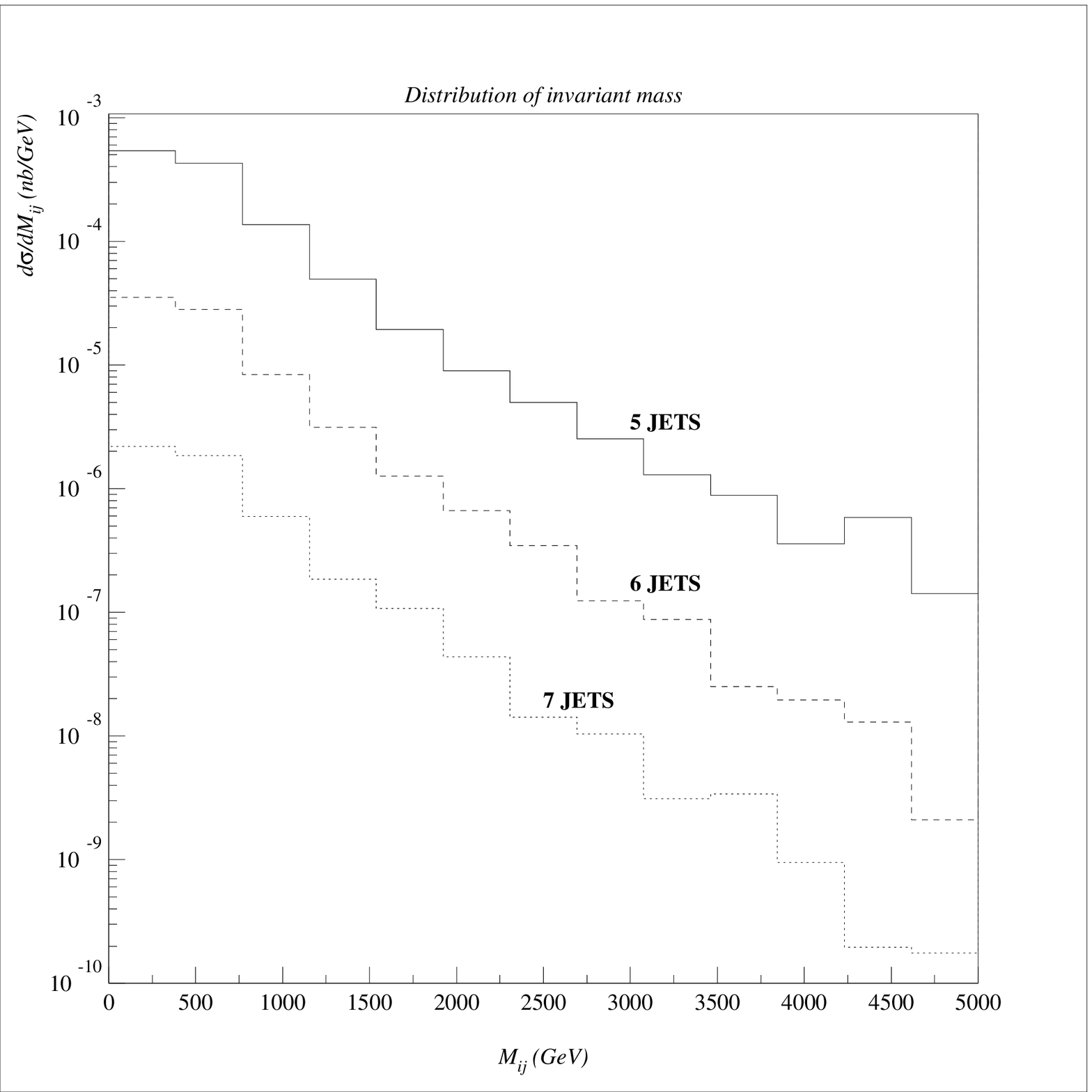,height=9cm,width=10cm}
\end{center}
\caption{Invariant mass distributions for 5,6 and 7 jet production}
\end{figure}

\newpage

\section*{Appendix A: The complete recursion relations }
Below we give the full recursion equations including the colour.
\begin{itemize}
\item Gluon field $A^\mu(p)$
\bqa
A^\mu_{AB}(P)&=& \frac{g_s}{2 P^2} \sum_{P=p_1+p_2}
\{ \bar{\psi}_B(p_1) \gamma^\mu \psi_A(p_2) - \frac{1}{3} 
\left( \sum_C \bar{\psi}_C(p_1) \gamma^\mu \psi_C(p_2) \right) \delta_{AB} \} \; \sigma(p_1,p_2) \nn \\
&+& \frac{g_s}{2 P^2} \sum_{P=p_1+p_2} V^{\mu \nu}_{\lambda}(P,p_1,p_2)
\{ A^{\nu}_{AC}(p_1) A^{\lambda}_{CB}(p_2)
  -A^{\lambda}_{AC}(p_2) A^{\nu}_{CB}(p_1) \} 
\; \sigma(p_1,p_2) \nn \\
&+& \frac{ig_s}{2 P^2} \sum_{P=p_1+p_2} X^{\mu \nu}_{\lambda \rho}
\{ A^{\nu}_{AC}(p_1) H^{\lambda \rho}_{CB}(p_2)
  -H^{\lambda \rho}_{AC}(p_2) A^{\nu}_{CB}(p_1) \} \; \sigma(p_1,p_2) \nn \\ \nn
\eqa
\item Auxiliary field $H_{\mu \nu}(p)$
\[
H^{\mu \nu}_{AB}(P)= \frac{(ig_s)}{4} \sum_{P=p_1+p_2} X^{\mu \nu}_{\lambda \rho}
\{ A^{\lambda}_{AC}(p_1) A^{\rho}_{CB}(p_2)
  -A^{\rho}_{AC}(p_2) A^{\lambda}_{CB}(p_1) \} \; \sigma(p_1,p_2) \nn
\]
\item Fermion field $\psi_A(p)$
\[
\psi_A(P)=\frac{g_s}{P^2} \sum_{P=p_1+p_2}P\cfeyn  
 A\cfeyn_{AB} (p_1) \;  \psi_B(p_2) \; \sigma(p_1,p_2) \nn
\]
\item Anti-fermion field $\bar{\psi}_A(p)$
\[
\bar{\psi}_A(P)=\frac{g_s}{P^2} \sum_{P=p_1+p_2} 
 \bar{\psi}_B(p_2) A\cfeyn_{BA} (p_1) P\cfeyn \;  \sigma(p_1,p_2) \nn
\]
\end{itemize}
%
%
\section*{Appendix B: Counting diagrams}
In this appendix we show how the number of tree-level graphs for a general
QCD process can be determined. This is done in the same  recursive manner
as that in which the amplitudes are computed.
Let there be $k$ quark flavours. We shall compute the number of graphs
that take a single specified parton into a number of specified partons,
as follows. Let us denote by 
$
a_i(n_0,n1,\bar n_1,n_2,\bar n_2,\ldots,n_k,\bar n_k)
$
the number of graphs with a quark of type $i$ coming in and ending up in
$n_0$ gluons, $n_1$ quarks of type 1, $\bar n_1$ antiquarks of type 1, and
so on. We define the following generating function:
\begin{eqnarray}
&&\psi(z,x_1,\bar x_1,x_2,\bar x_2,\ldots,x_k,\bar x_k)
=\nonumber\\&& \sum\limits_{n_0,n_1,\ldots,\bar n_k\ge0}
a_i(n_0,n_1,\bar n_1,n_2,\bar n_2,\ldots,n_k,\bar n_k)
{z^{n_0}x_1^{n_1}\bar x_1^{\bar n_1}x_2^{n_2}\cdots \bar x_k^{\bar n_k}
\over n_0!n_1!\bar n_1!n2!\cdots\bar n_k!}\;\;.  \nn
\end{eqnarray}
The generating functions $\bar{\psi}$ for incoming antiquarks, and $\phi$
for incoming gluons, are defined in an analogous manner.
The Feynman rules of QCD then tell us that the various generating functions
are related in the following manner:
\begin{eqnarray}
\psi_i &=& x_i + \psi_i\phi\;\;,\nonumber\\
\bar{\psi}_i &=& \bar x_i + \bar{\psi}_i\phi\;\;,\nonumber\\
\phi &=& z + {1\over2}\phi^2 + {1\over6}\phi^3 +
\sum\limits_{i=1}^k\psi_i\bar{\psi}_i\;\;. \nn
\end{eqnarray}
These can trivially be solved in such a way that everything is expressed in
terms of $\phi$:
\[
\psi_i = x_i/(1-\phi)\;\;\;,\;\;\;\bar{\psi}_i = \bar x_i/(1-\phi)\;\;\;, \nn
\]
and
\[
\phi = z + {1\over2}\phi^2 + {1\over6}\phi^3 + {\xi\over(1-\phi)^2}\;\;\;,\;\;\;
\xi = \sum\limits_{i=1}^kx_i\bar x_i\;\;.  \nn
\]
The number of different quark flavours is seen not to be a complication here.
By computer algebra, the function $\phi$ can easily be obtained to quite
high order in $z$ and the $x$'s, and the number of diagrams read off
from the corresponding coefficient.
A similar discussion can be found in \cite{manganoetal}, but we feel that
the above procedure is simpler and moreover leads itself to estimates of the
asymptotic number of graphs for very large multiplicities
\cite{diagramcountingR&P}.
\section*{Appendix C: Counting distinct processes}
We start by tabulating the five types of initial states (not counting
trivial charge conjugation), 
with the corresponding
number of distinct processes, and the functions needed to compute their
multiplicity factors. We denote by $n$ the number 
of final state partons:

\begin{center}
\begin{tabular}{|lc|c|c|}
\hline
\multicolumn{2}{|c|}{initial-state type} 
& distinct processes & multiplicity factor
\\
\hline
A &($gg$) & $C_1(n)$ & $\chi(n_0,n_1,\ldots,n_f;f)$
\\
\hline
B &($q\bar{q}$) & $C_2(n)$  & $\chi(n_0,n_2,\ldots,n_f;f-1)$
\\
\hline
C &($gq$ and $qg$) & $C_2(n-1)$ & $\chi(n_0,n_2,\ldots,n_f;f-1)$
\\
\hline 
D &($qq$) & $C_2(n-2)$  & $\chi(n_0,n_2,\ldots,n_f;f-1)$
\\
\hline
E &($q q^\prime$ and $ q\bar{q}^\prime$)  & $C_3(n-2)$
& $\chi(n_0,n_3,\ldots,n_f;f-2)$
\\
\hline
\end{tabular} 
\end{center}

In order to clarify what we mean we consider the example of the
type A initial state. Each distinct process is defined by an array  
$(n_0,n_1,\ldots,n_f)$. For instance, in the case of 
four-jet production we have

\begin{center}\begin{tabular}{cc}
(4,0,0,0,0,0) & $gg\to gggg$
\\
(2,1,0,0,0,0) & $gg\to gg q\bar{q}$
\\
(0,2,0,0,0,0) & $gg\to q\bar{q} q\bar{q}$
\\
(0,1,1,0,0,0) & $gg\to q\bar{q} r\bar{r}$
\\
\end{tabular}\end{center}

Therefore, in order to count the distinct processes 
we need the following three functions:

\[
C_1(n)=\sum_{n_0+2n_1+\ldots+2n_f=n} \Theta(n_1\ge n_2\ge\ldots\ge n_f) \nn
\]
\[
C_2(n)=\sum_{n_0+2n_1+\ldots+2n_f=n} \Theta(n_2\ge n_3\ge\ldots\ge n_f)  \nn
\]
and
\[
C_3(n)=\sum_{n_0+2n_1+\ldots+2n_f=n} \Theta(n_3\ge n_4\ge\ldots\ge n_f)  \nn
\]

Of course each distinct process, given by the array $(n_0,n_1,\ldots,n_f)$
has a multiplicity factor that is easily computed:

\[
\chi(n_0,n_1,\ldots,n_f;f)=n_f(n_f-1)...(n_f-j+1)/j!  \nn
\]
where $j$ is defined as
\bqa
j=f    &\mbox{if}& \prod_{i=1}^{f}{n_i}\ne 0    \nn \\
j=f-1  &\mbox{if}& \prod_{i=1}^{f-1}{n_i}\ne 0  \nn  \\
\ldots \nn \\
j=1  &\mbox{if}& n_1\ne 0  \nn \\
j=0  &\mbox{otherwise}&  \nn
\eqa

Results for $f=5$ final state flavours, 
and $f=4$ initial state flavours, are shown in the table of section 2, in the
discussion of flavour treatment, just to compare with ref.~\cite{Kuijf}.

Moreover it is very easy to produce the list of the distinct processes
to be computed for each case as well as the multiplicity factors $\chi$.
A code doing this is available.

To study high-$n$ behaviour we may use the generating function technique.
Then we get
\bqa
F_1(x)&=&\sum_{n=0}^{\infty} C_1(n) x^n = \frac{1}{(1-x)}
\prod_{j=1}^{f} \frac{1}{(1-x^{2j})}
\nn \\
F_2(x)&=&\sum_{n=0}^{\infty} C_2(n) x^n = 
\frac{1}{(1-x)(1-x^2)}\prod_{j=1}^{f-1} \frac{1}{(1-x^{2j})}
\nn \\
F_3(x)&=&\sum_{n=0}^{\infty} C_3(n) x^n = 
\frac{1}{(1-x)(1-x^2)^2}\prod_{j=1}^{f-2}  \frac{1}{(1-x^{2j})} \nn
\eqa
As one can easily see, that the order of the pole at $x=1$ is always
given by $f+1$. 

Our results may be compared directly to
those obtained in ref.~\cite{Kuijf}. In fact, using our method, 
we were able to detect
several errors in Table 9.3 page 125, among which the fact that a processes is 
missing for the cases $m=6$ and $m=7$, namely 
$qr\to qrr\bar{r}(g)$. Correcting for this error we get full agreement.

\end{document}